\documentclass[aps,prl,footinbib,floats,twocolumn,showpacs,superscriptaddress,longbibliography]{revtex4-1}

\usepackage[gen]{eurosym}
\usepackage{natbib}
\usepackage[utf8]{inputenc}
\usepackage{amsmath, amsthm, amssymb}
\usepackage{enumitem}
\usepackage{graphics,graphicx}
\usepackage{times}
\usepackage{dcolumn}
\usepackage{hyperref}
\usepackage{color}

\usepackage[all]{xy}
\usepackage{mdwlist}

\renewcommand{\(}{\left(}
\renewcommand{\)}{\right)}

\begin{document}

\renewcommand*\thesection{\arabic{section}}
\newcommand{\beq}{\begin{equation}}
\newcommand{\eeq}{\end{equation}}
\newcommand{\sss}{\scriptscriptstyle}

\title{Decongestion of urban areas with hotspot-pricing}

\author{Albert Sol\'e-Ribalta}
\affiliation{Departament d'Enginyeria Inform\`atica i Matem\`atiques,
Universitat Rovira i Virgili, 43007 Tarragona, Catalonia, Spain}
\affiliation{Internet Interdisciplinary Institute,
Universitat Oberta de Catalunya, 08018 Barcelona, Catalonia, Spain.}

\author{Sergio G\'omez}
\affiliation{Departament d'Enginyeria Inform\`atica i Matem\`atiques,
Universitat Rovira i Virgili, 43007 Tarragona, Catalonia, Spain}

\author{Alex Arenas}
\affiliation{Departament d'Enginyeria Inform\`atica i Matem\`atiques,
Universitat Rovira i Virgili, 43007 Tarragona, Catalonia, Spain}
\affiliation{IPHES, Institut Catal\`a de Paleoecologia Humana i Evoluci\'o Social,
43003 Tarragona, Catalonia, Spain}

\begin{abstract}
The rapid growth of population in urban areas is jeopardizing the mobility and air quality worldwide. One of the most notable problems arising is that of traffic congestion which in turn affects air pollution. With the advent of technologies able to sense real-time data about cities, and its public distribution for analysis, we are in place to forecast scenarios valuable to ameliorate and control congestion. Here, we analyze a local congestion pricing scheme, hotspot pricing, that surcharges vehicles traversing congested junctions. The proposed tax is computed from the estimation of the evolution of congestion at local level, and the expected response of users to the tax (elasticity). Results on cities' road networks, considering real-traffic data, show that the proposed hotspot pricing scheme would be more effective than current mechanisms to decongest urban areas, and paves the way towards sustainable congestion in urban areas.
\end{abstract}

\maketitle


\section{Introduction}\label{intro}
Urban life is characterized by a huge mobility, mainly motorized. Amidst the complex urban management problems there is a prevalent one: traffic congestion. INRIX Traffic Scorecard (\url{http://www.inrix.com/}) reports the rankings of the most congested countries worldwide in 2014. US, Canada and most of the European countries are in the top 15, with averages that range from 14 to 50 hours per year wasted in congestion, with their corresponding economic and environmental negative consequences. Several approaches exist to efficiently design road networks \cite{Yang1998,szeto2015sustainable} and routing strategies \cite{Bast2007,qian2013hybrid}, however, the establishment of collective actions to prevent or ameliorate urban traffic congestion require further improvements, given the complex behavior of drivers.

An striking, as well as controversial, strategy to address the problem is congestion pricing \cite{BoarnetBriefImpactRoadPricing2014,de2004congestion,friesz2004dynamic}. It consists in taxing vehicles for accessing a road/area, at certain times, based on the supply-demand model \cite{samuelson1996economia}. Since the supply quantity is fixed (no more lanes or roads are usually added to the transportation network) the access to demanded areas is taxed. Two main types of congestion pricing \cite{dePalma20111377} exist: i) road pricing, where vehicles are charged for using a particular road section ---such as freeways, ring roads, tunnels or bridges---, and ii) cordon pricing, where vehicles are charged to access a particular zone susceptible to traffic congestion ---such as historical towns, business districts or simply crowded areas---. A similar variant is area pricing, where the tax applies per day. While road pricing is usually understood as a Pigovian tax to compensate for the externalities caused by drivers  \cite{arnott1994EconomicsTrafficCongestion}, cordon pricing can be understood solely as an incentive for reducing the traffic congestion and improving the air quality of the city \cite{Parrish2009CleanAirMegacities}, but eventually also becomes a tax income for urban areas.

Generally speaking, cordon/area pricing is, in general, effective in reducing the overall amount of cars accessing restricted areas and reducing pollution \cite{Moroni2013AirQualityAreaCMilan,Besser2016ImpactOfStockholmCongCharge} but it is still insufficient to reduce congestion hotspots within the taxed zone. These hotspots usually correspond to junctions and are problematic for the efficiency of the network as well as for the health of pedestrians and drivers. It has been shown \cite{Petersson1978ExposureToTrafficExhaust} that drivers in-queue are the most affected collective to car exhaust pollution inhalation. In addition, these hotspots are usually located in the city center, magnifying the problem \cite{Raducan2009PolutansBucarest}. Assuming that congestion is an inevitable consequence of urban motorized areas, the challenge is to develop strategies towards a sustainable congestion regime at which delays and pollution are under control.

Since ten years ago the scientific community has proposed models to analyze the problem of traffic congestion \cite{tadic04,liang05,gawron1998iterative,nagel2008multi} and decongestion \cite{singh05,yan1996optimal,arnott1993structural}, pollution generated by traffic \cite{kickhofer2016towards,grote2016including,misra2013integrated,panis2006modelling}, transitions between traffic states \cite{guimera2002dynamical,echenique2005dynamics,kim09}, and the design of optimal topologies \cite{donetti05,danila06,bart06,li10} and algorithms \cite{ramasco10,sce10} to avoid it. The focus of attention of most of the previous works was the onset of congestion, which corresponds to a critical point in a phase transition, and how it depends on the topology of the network and the routing strategies used. However, the proper analysis of the system after congestion has remained analytically slippery. It is known that when a transportation network reaches congestion, the travel time and the amount of vehicles queued in a junction diverge \cite{doro08}.

Here, we rely on our Microscopic Congestion Model (MCM) to identify urban traffic hotspots in real scenarios and devise a mechanism to palliate its congestion \cite{Sole-RibaltaMCM}. The mechanism is a taxing scheme that charges directly vehicles crossing congested spots (junctions) considering the overall topological structure and traffic functionality of the network. The aim is to eliminate the congestion hotspots using a network topology pay-per-use scheme. Specifically, we build up a flow model based on two steps: (1) detection of the hotspots using MCM, and (2) prediction of the required tax to be applied to every congested junction to encourage drivers to divert the excess flow to neighboring and less congested regions. Our approach follows a similar idea to the one proposed by Vickrey back in 1963 \cite{vickrey1963pricing}, with the main difference that we now can analytically predict the model behaviour considering real data.

\section{Microscopic Congestion Model for hotspot detection} \label{MCM}

The taxing scheme we propose relies on the identification of the city hotspots. Although stochastic micro simulations could be used to obtain the required parameters, here we focus on a recently developed analytically tractable model, called {\em Microscopic Congestion Model} (MCM) \cite{Sole-RibaltaMCM}. The model assumes the following discrete and stylised car-flowing dynamics. At each time step vehicles are injected into each junction $i$ of the system by exogenous process at rate $\rho_i$ following a given distribution. During the following time steps, vehicles navigate towards their destination following the shortest-path; the model can also consider other traffic dynamics, such as diffusion dynamics or random walks, or even other mobility models such as the gravitational model \cite{zipf1946p,de1994modelling} or the radiation model \cite{simini2012universal,ren2014predicting}. To simulate the waiting time of vehicles at road junctions, we assign a first-in-first-out queue to each one\footnote{Alternatively, a more realistic approach would be to attach queues to links instead of junctions, and modify the equations accordingly. Here, we just rely in MCM as described in \cite{Sole-RibaltaMCM}.}. We suppose these queues have a maximum processing rate, $\tau_i$, that mimics the physical constraints of the junction. That is, at most $\tau_i$ vehicles cross a junction per unit time. Similar car-flowing dynamics, also based on queues, have been previously used to develop agent-based models for traffic analysis \cite{gawron1998iterative,cetin2003large,raney2003agent,grether2012simulation} and pricing policies \cite{yan1996optimal,kickhofer2016towards}. As it is shown in \cite{Sole-RibaltaMCM} the previous scenario has a critical generation rate $\rho_c$ such that, for any generation rate larger than $\rho_c$, the network is not able to route or absorb all incoming vehicles. In this situation, the total amount of vehicles $Q(t)$ in the network grows proportionally to time. Locally, each junction of the network has its own critical injection rate $\rho_{c_i}$ which is governed by its node effective betweenness $B_i$ \cite{newman2010networks}, $\rho_{c_i} \propto 1/{B_i}$. The first junction to reach congestion defines the network critical injection rate, $\rho_c$.

\begin{figure*}[tb]
	\begin{center}
		\includegraphics[width=0.95\textwidth]{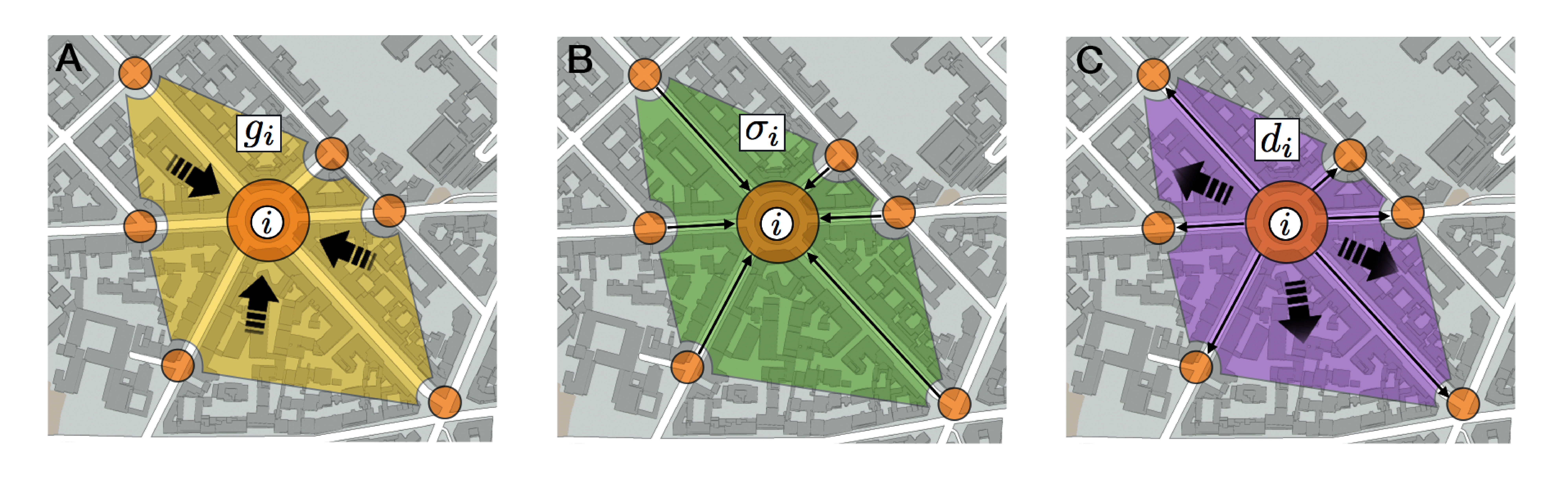}
	\end{center}
        \caption{Illustration of the variables of the MCM model. {\bf (A)} Vehicles entering junction $i$ from the area surrounding $i$. {\bf (B)}  Vehicles entering junction $i$ from its neighboring junctions. {\bf (C)} Vehicles leaving junction $i$, either to go to other neighboring junctions or to finishing the trip in its surrounding area.}
        \label{fig:modelExplanation}
\end{figure*}

The MCM describes the full state of system for any amount of congested junctions. The MCM is based on assuming that the growth of vehicles observed in each congested node of the network is constant, which corresponds to the stationary state. This assumption allows us to describe, with a set of balance equations (one for each node), the increment of vehicles in the junction queues'. Mathematically, the increment of the vehicles per unit time at every junction $i$ of the city, $\Delta{q}_{i}$, satisfies:
\begin{equation}
	\Delta{q}_{i} = g_{i} + \sigma_{i} - d_{i},
	\label{BE}
\end{equation}
where $g_{i}$ is the average number of vehicles entering junction $i$ from the area surrounding $i$, $\sigma_{i}$ is the average number of vehicles that arrive to junction $i$ from the adjacent links of that junction, and $d_{i} \in [0,\tau_i]$ corresponds to the average of vehicles that actually finish in junction $i$ or traverse towards other junctions. A graphical explanation of the variables of the model is shown in Fig.~\ref{fig:modelExplanation}. The system of eqs.~(\ref{BE}) defined for every node $i$, is coupled through the incoming flux variables $\sigma_{i}$, that can be expressed as
\begin{equation}
	\sigma_{i} = \sum^{S}_{j=1} P_{ji} p_{j} d_{j},
	\label{sigma}
\end{equation}
where $P_{ji}$ accounts for the routing strategy of the vehicles (probability of going from $j$ to $i$), $p_j$ stands for the probability of traversing junction $j$ but not finishing at $j$ and $S$ is the number of nodes in the network.

For each junction $i$, the onset of congestion is determined by $d_i=\tau_i$, meaning that the junction is behaving at its maximum capability. Thus, for any flux generation rate ($g_i$), routing strategy ($P_{ij}$) and origin-destination probability distribution, eqs.~(\ref{BE}) can be solved using an iterative approach to predict the increase of vehicles per unit time at each junction of the network ($\Delta{q}_{i}$). See \cite{Sole-RibaltaMCM} for further details of the model and a detailed description on how to obtain the system variables.

In the following sections, we apply the {\it Hotspot Pricing} scheme in cities that are in the congested regime, $\Delta q_i(t) >0$ for some junctions. We use the MCM to obtain the data-driven state of the system. The very basic idea of the hotspot pricing scheme is to reduce the excess of vehicles that accumulate at the queues of each congested junction to reach to the desired level.

In this work, we have used two real source and destination distributions, obtained through Open Data portals, that consider the ingoing and outgoing flux of vehicles of the cities of Milan (Italy) and Madrid (Spain). Although different origin and destination models can be used (e.g. gravity and radiation models), here we assume a ``Home-to-Work'' travel pattern, where vehicles arrive from the outskirts of the city and go to the city center. Consequently, traffic is generated at rate $\rho_i$ in the peripheral junctions of the network (arrival to the city), go to a randomly selected junction (arrival to work) and then returns back to a peripheral junction (return home). We do not consider trips with origin and destination inside the city center since public transportation systems (e.g., train or subway) usually constitute a better alternative than private vehicles for those trips.

\section{Hotspot pricing scheme} \label{HSP}

To reduce congestion levels, we propose to tax the junctions where $\Delta{q}_{i}~>~0$. Clearly, the higher the tax, the fewer the drivers that will want to pass through the taxed junction and consequently the lower the congestion. To estimate the required tax for each junction, we use the economic concept of {\bf elasticity} \cite{frank2007microeconomics}. The elasticity measures the response of the demand of a good in terms of an increase of its price and it is formally obtained as the ratio between the relative increase of the demand of a good and the relative increase of its price. The elasticity has been successfully used to predict the electricity demand given an increase of its price \cite{labandeira2012estimation}, to forecast fuel consumption \cite{dahl2012measuring}, to price in the Internet transit market \cite{valancius2011many} or to obtain airport charges given their passengers profiles \cite{pels2004economics} and, within the context of transportation planning, to measure the effect of an increase of fares on the public transport demand \cite{paulley2006demand} or to model the effect and consequences of toll roads \cite{olszewski2005modelling,swan2010empirical}. Here, we use the elasticity the other way around. Instead of predicting the traffic demand that we would observe given a tax, we predict the tax to obtain the desired reduction in the congestion.
The elasticities of road taxing and cordon pricing are negative, meaning that an increase of tax produces a decrease of the traffic \cite{Litman2012ElasticitiesOfTransportDemand,BoarnetBriefImpactRoadPricing2014}. These elasticities lay between $-0.2$ to $-0.9$ for cordon pricing schemes and between $-0.03$ to $-0.5$ for road tolls, and they depend on the country and on the application. The lower the elasticity the less reactive is the society towards taxing schemes. Thus, the demand curve with respect to price follows a power law function which can be fitted given an observation and a slope (the elasticity). This curve can be used to predict how traffic is affected by a change of tax. In the rest of the article, we assume that an equivalent tax produces the same effect on the incoming flow at each junction. Note that this assumption has been made because of a lack of actual information about the real elasticities, although this does not undermine the essential behavior of the model that can be fitted with observed elasticities when available.

The predicted fraction of flow of vehicles after a tax $c$ is applied (i.e. those vehicles that decide to pay instead of diverting their paths) is given by:
\begin{equation}
	\phi = \phi_0\left(\frac{c}{c_0}\right)^{\mu}
	\label{eq:phi}
\end{equation}
where $\phi_0$ is the observed fraction of flow after applying the tax $c_0$, and $\mu$ is the elasticity value (see \cite{dePalma20111377} for a detailed description of the technological implementation of taxes).

To approach zero congestion, the proposed hotspot pricing scheme consists in taxing each congested junction $i$ (that is, junctions with $\Delta{q}_{i} > 0$) to eliminate the accumulation of vehicles,
\begin{equation}\label{}
  c_i=c_{0} \left(\frac{\phi_i}{\phi_0}\right) ^{1/\mu}
  = c_{0} \left(1-\frac{\Delta q_i}{\rho_i + \sigma_i}\right)^{1/\mu}.
  \label{eq:tax}
\end{equation}
where $\sigma_i$ is the total amount of vehicles arriving at junction $i$ per unit time coming from the neighbouring junctions. The term between parentheses represents the maximum flux, with respect to the original incoming flux, the junction can deal with without being congested.

\section{An application of the model to the traffic in Milan}

\begin{figure}[tb]
	\centering
	\includegraphics[width=0.95\columnwidth]{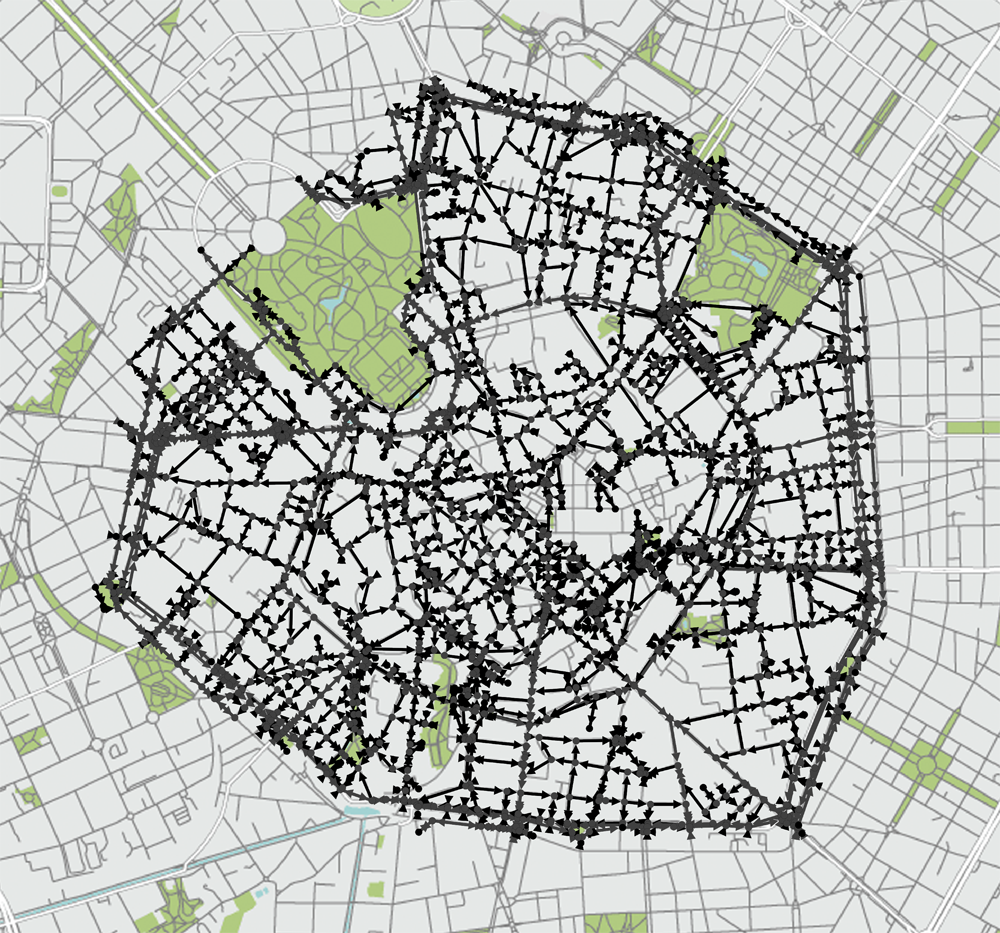}
	\caption{Topology of the road network inside the Area-C of Milan. Data gathered from Open Street Map (\url{http://www.openstreetmap.org}).}
	\label{fig:milano_topology}
\end{figure}

We have analyzed the potential effect of the proposed hotspot pricing scheme using the predictions of our simulations in the city with largest INRIX value, Milan (Italy). Milan actually applies a cordon pricing scheme to reduce the transit of vehicles inside the historical city center; they call ``Area-C'' to this restricted traffic area. Their taxing scheme is monitored by camcorders at 43 gates. The individual tax applied depends on the type of car and also on its activity. The tax ranges from free tax for electric vehicles, scooters and public transport to 5\euro{} for non-resident vehicles. Results published by ``Comune di Milano'' show that, after the cordon charge establishment, there has been a reduction of approximately 45\% of non-resident vehicles and around 35\% of the total traffic. This observation allows to assimilate the tax value $c_0 = 3\euro{}$ corresponding to an average reduction of $\phi_0 = 0.35$. We recover the original flux of vehicles, previously to the establishment of the Area-C, rescaling the observed flux considering the observed reduction. With respect to the value of the elasticity, we cannot predict precisely how responsive will be vehicle users to the hotspot pricing (the elasticity) but current elasticities observed for cordon pricing and toll roads should be closely related in sign (negative) and in magnitude. In the following, we assume an elasticity of $\mu=-0.1$ which is compatible with the values observed for road tolls in other cities.

To apply the hotspot pricing scheme, we first gather data about the road network topology using Open Street Map (OSM), a well-known data source for traffic analysis \cite{behrisch2011sumo,sommer2011bidirectionally}. OSM data represents each road (or way) with an ordered list of nodes which can either be road junctions or simply changes of the direction of the road. We have obtained the required abstraction of the road network building a simplified version of the OSM data which only accounts for road junctions (nodes). Then, for each pair of adjacent junctions we have queried the real travel distance (i.e.\ following the road path) using the API provided by Google Maps. The resulting network corresponds to a spatial weighed directed network \cite{barth2011} where the driving directions are represented and the weight of each link indicates the expected traveling time between two adjacent junctions (see Fig.~\ref{fig:milano_topology}). We build up the dynamics of the model analyzing real traffic data provided by Telecom Italia for their Big Data Challenge. The data provides, for every car entering the cordon pricing zone in Milan during November and December 2013, an encoding of the car's plate number, time and gate of entrance. This allows us to obtain the (hourly) average incoming and outgoing traffic flow, for each gate of the cordon taxed area. Without any extra information, we are forced to consider all vehicles of the same type, and we assume that all are required to pay the same type of tax. Given the previous topology and traffic information, we first compute the expected traffic within the city (see Sect.~\ref{MCM}), and then calculate the required tax (see Sect.~\ref{HSP}).

\subsection{The cost of zero congestion}

\begin{figure*}[tb]
	 \begin{tabular}{lll}
		{\bf A} & {\bf B} & {\bf C}\\
		\includegraphics[width=.32\textwidth]{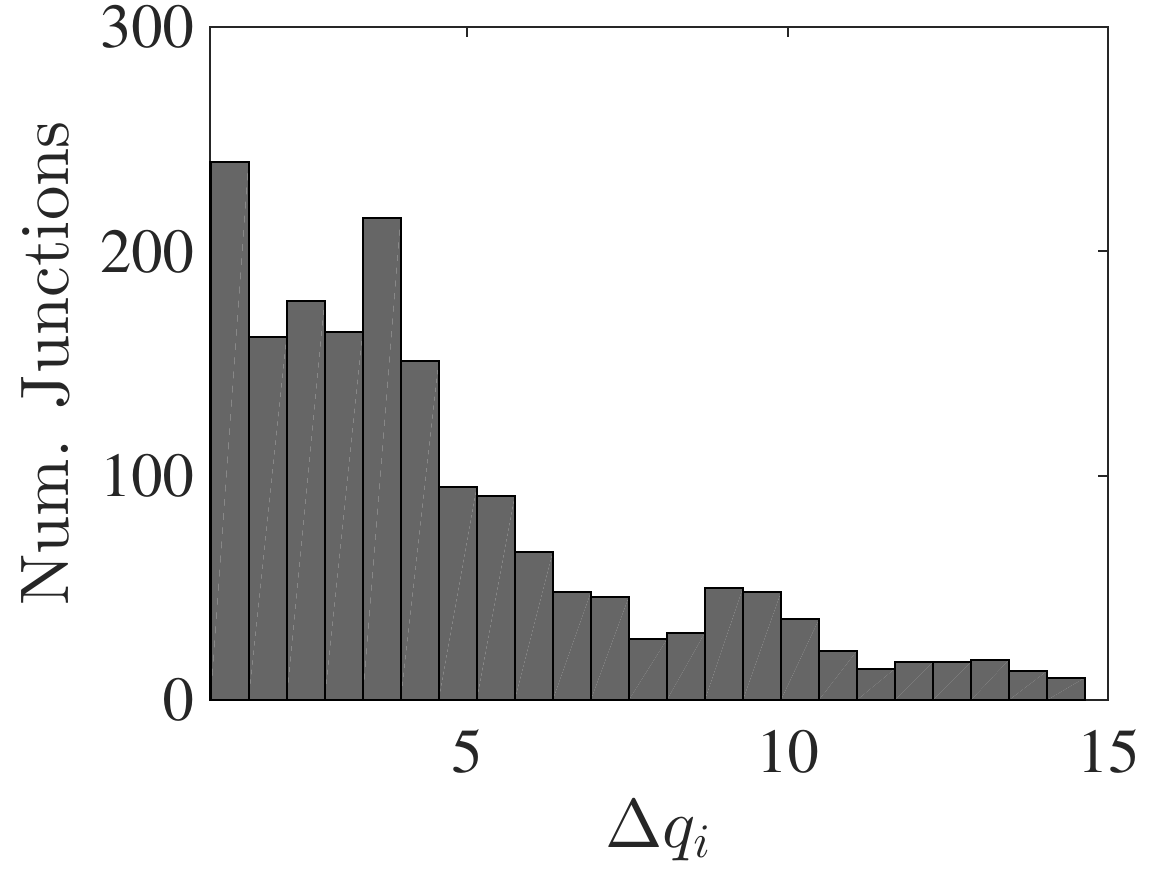} &
		\includegraphics[width=.32\textwidth]{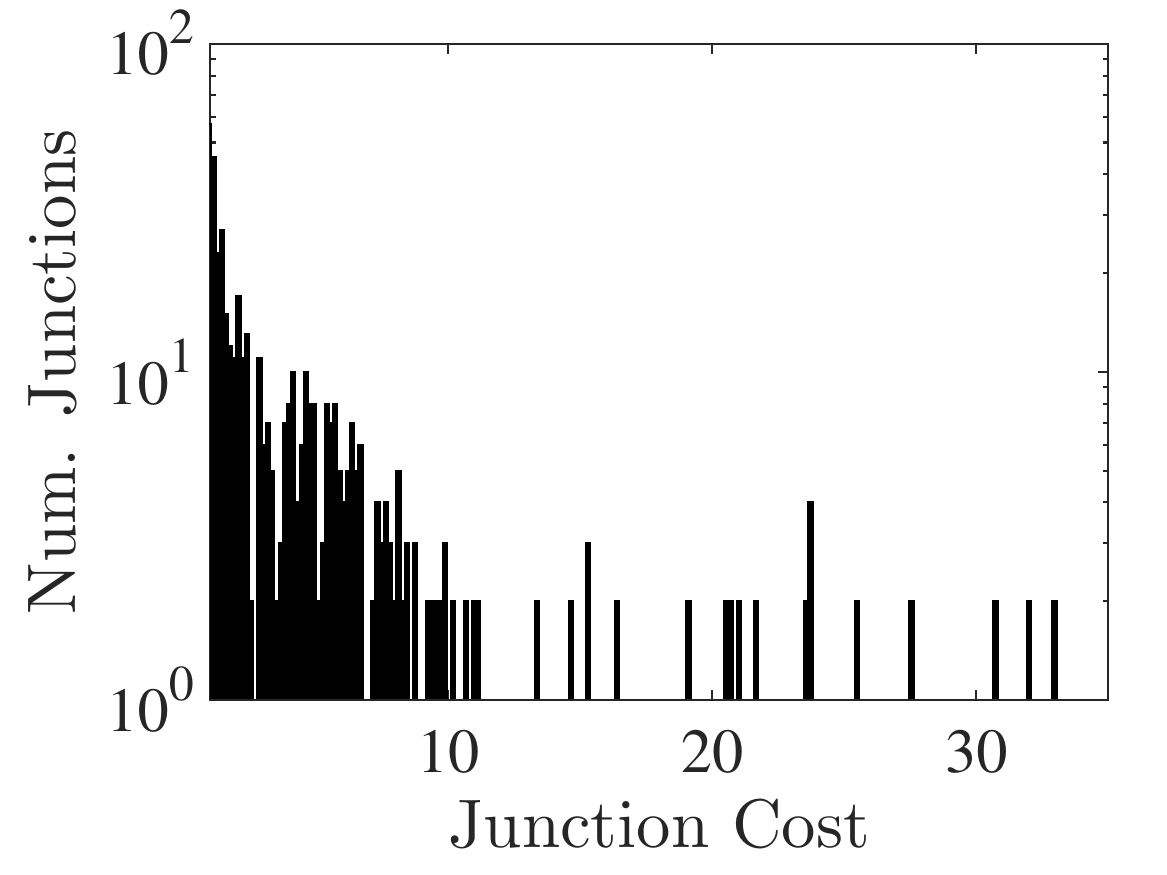} &
		\includegraphics[width=.32\textwidth]{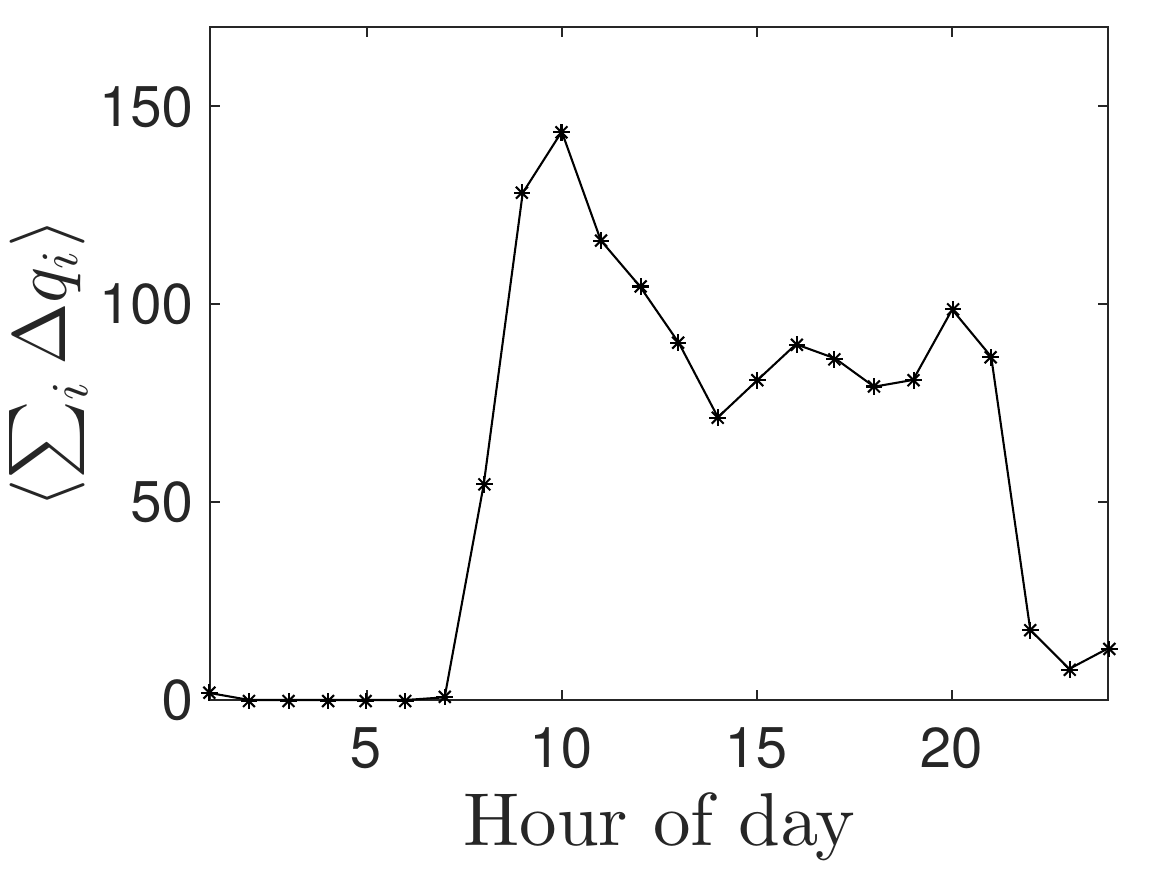}
 	 \end{tabular}
	\caption{Congestion and hotspot pricing for zero congestion in Milan. ({\bf A}) Distribution of the predicted congestion, $\Delta q_i$, of a week. ({\bf B}) Distribution of the required junction prices (in euros) to eliminate the congestion of a week. ({\bf C}) Average predicted congestion per hour of the day (in vehicles per minute).}
	\label{fig:removeAllCongestionMilan}
\end{figure*}

\begin{figure*}[htb]
	\centering
	\includegraphics[width=0.9\textwidth]{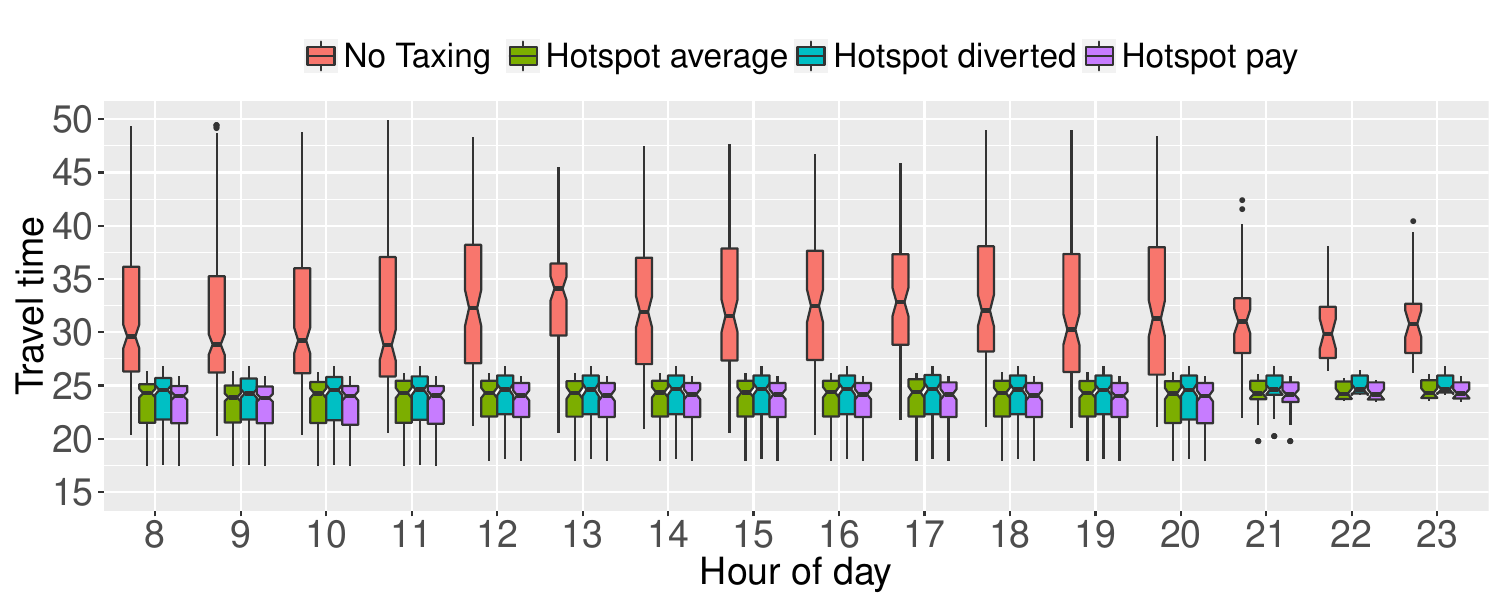}
	\caption{Comparison of the expected travel times (in minutes) of the trips affected by congested junctions before and after the zero congestion hotspot taxing scheme is applied, in Milan. For the hotspot pricing model, we show the distribution of travel times of trips crossing congested junctions after the hotspot establishment (labeled `Hotspot pay'), of the trips that now avoid the hotspots ('Hotspot diverted'), and of both of them together (`Hotspot average').}
	\label{fig:times_zero_congestion}
\end{figure*}

For illustration purposes, we first analyze the predicted cost to remove all congestion within the Area-C of Milan. To this aim, we have gathered for each hour of the day and each day of the week the ingoing flow of vehicles. Given the values of $d_i$, $\sigma_i$ and $\Delta q_i$, obtained using Monte Carlo simulations, we have computed the junction taxes to redistribute the required vehicles to achieve $\Delta q_i = 0$ at all junctions $i$. Figure~\ref{fig:removeAllCongestionMilan} reports the obtained results applying the double-step process described above only once. We see that most of the junctions have an increment, per unit time, $\Delta q_i$, below $5$ vehicles per minute which yields in general to a price per traversal below~$10$\euro{}. Extending the analysis to the full year, the annual income predicted by our model is 145M\euro{} in the case of taxing from $7$~a.m. to $7$~p.m., as it is done at present, or an income of 167M\euro{} in case of extending the taxing scheme to the full day. Note that even though the distribution is heavy tailed, the price at a set of junctions is almost prohibitive (around 25 \euro{}).

Figure~\ref{fig:times_zero_congestion} shows the reduction of travel times when the zero congestion hotspot taxing scheme is applied. Here we have supposed equal waiting times at all intersections except at the congested ones, where waiting times take into account the size of the queues. It is remarkable that the total travel times of the vehicles that cross and pay at the hotspots are only slightly better than the times for those vehicles avoiding the hotspots. This also evidences that the excess time for a driver not paying the tax is not substantial, being around a minute.

\subsection{Comparison with cordon pricing}

\begin{figure*}[tb]
	 \begin{tabular}{ll}
		{\bf A} \\
		\multicolumn{2}{c}{\includegraphics[width=0.80\textwidth]{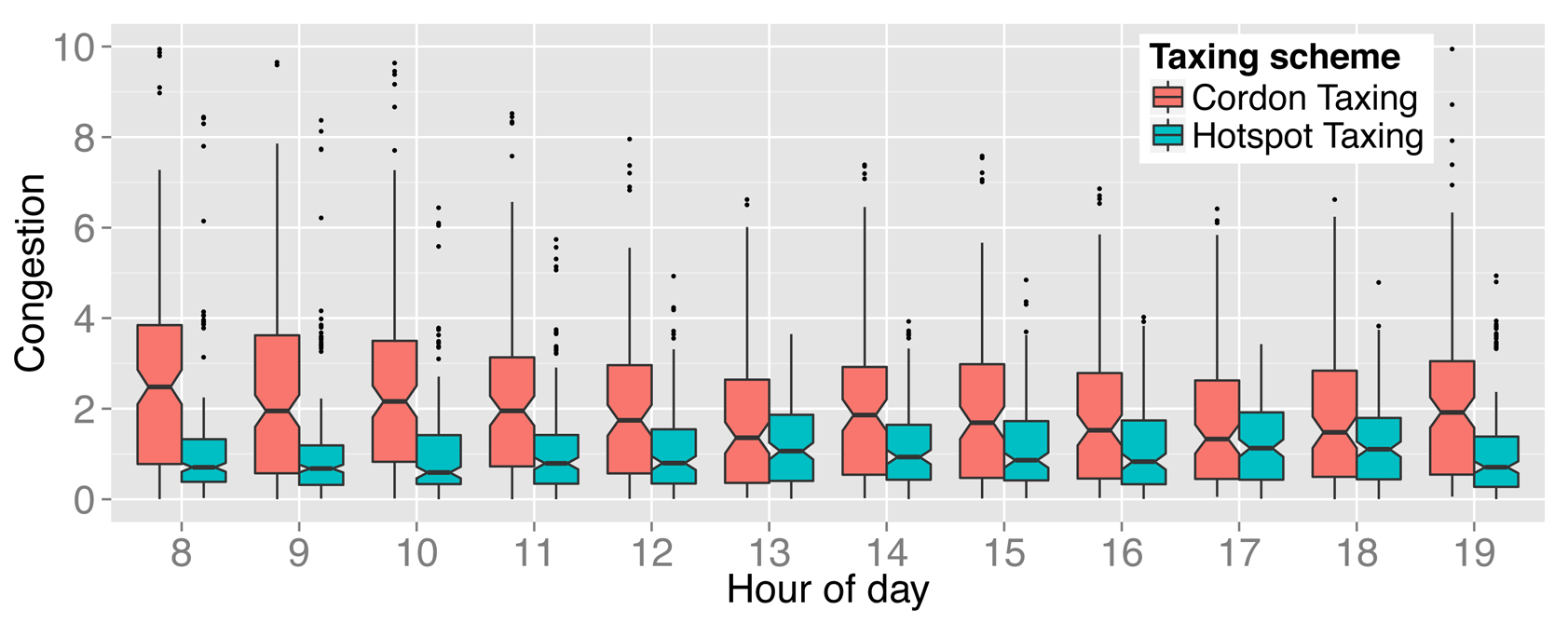}}\\
		{\bf B} & {\bf C} \\
		\includegraphics[width=0.45\textwidth]{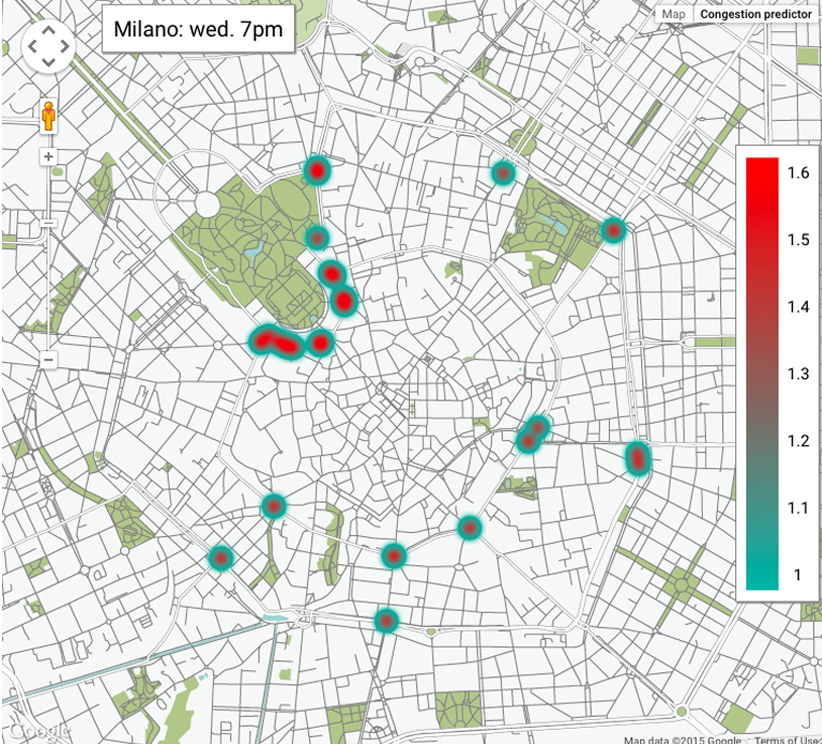} &
		\includegraphics[width=0.45\textwidth]{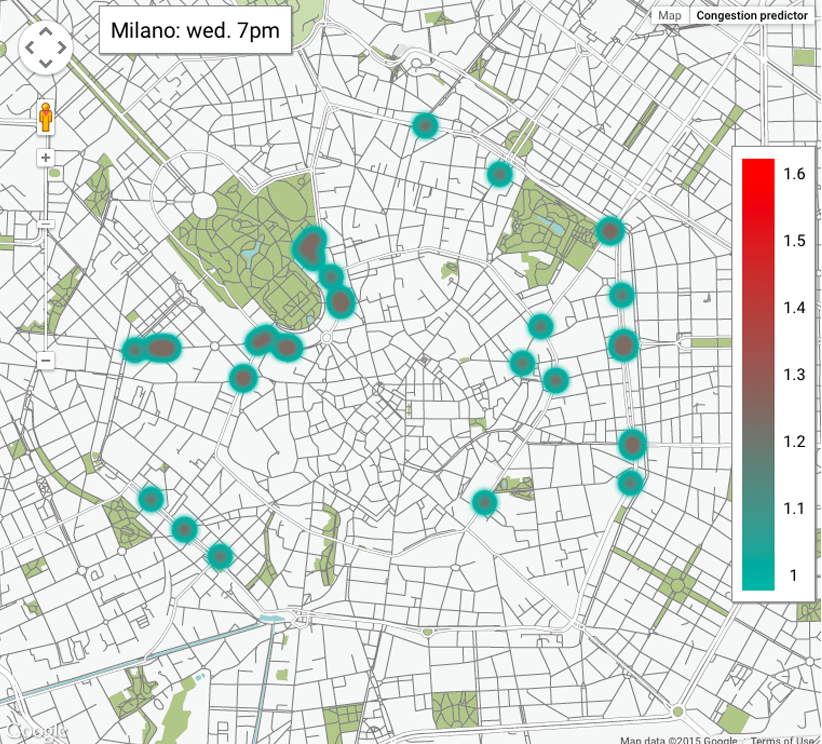}
 	 \end{tabular}
	\caption{({\bf A}) Distribution of the congestion after applying the different taxing schemes in the city of Milan. The value of the congestion is given by the accumulation rate of vehicles ($\Delta q_i$), in vehicles per minute, for the different congested junctions. The distributions are computed considering all data from Monday to Friday and they are shown grouped by hour of the day. Below, maps of Milan showing the expected ratios between incoming and outgoing vehicles of each junction after the establishment of the cordon pricing tax ({\bf B}) and the hotspot pricing scheme ({\bf C}). Junctions with a ratio greater than $1$ are congested since they receive more cars that the ones they can route.}
	\label{fig:congDiffusionMilan}
\end{figure*}

\begin{figure*}[htb]
	\centering
	\includegraphics[width=0.9\textwidth]{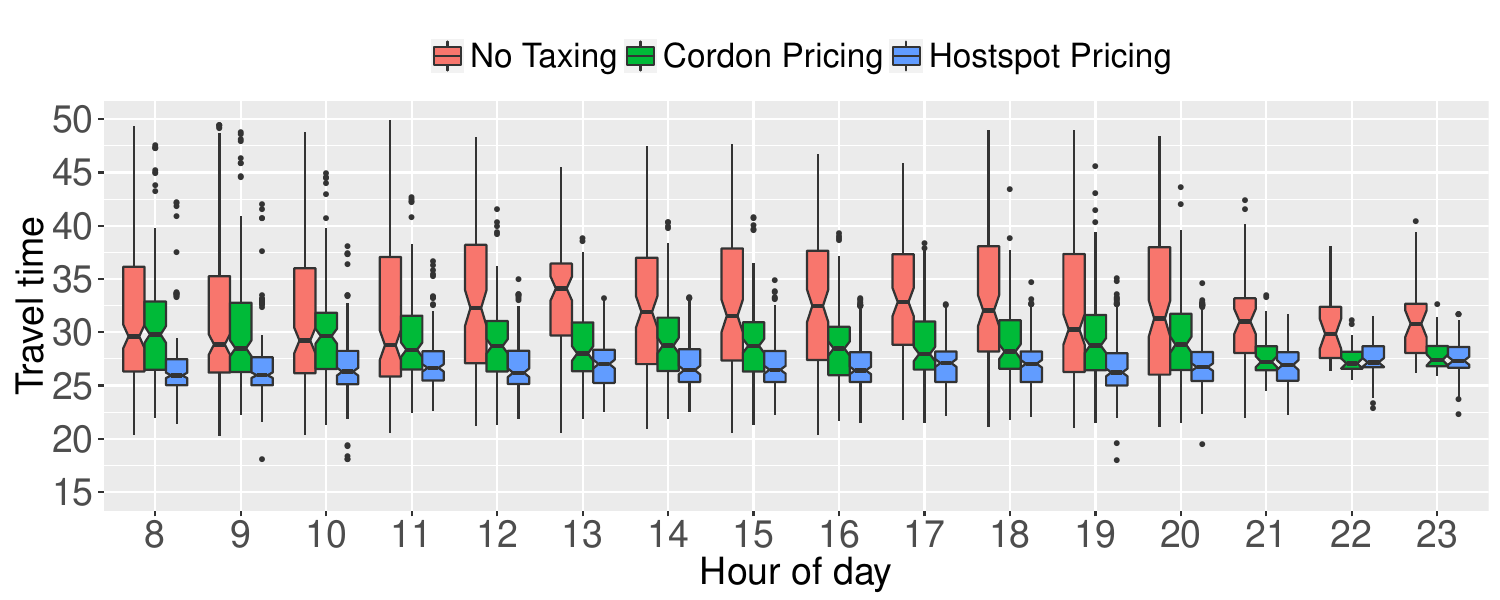}
	\caption{Comparison of travel times (in minutes) between no taxing, cordon pricing and hotspot pricing, with fixed total income, for the city of Milan.}
	\label{fig:times_same_revenue}
\end{figure*}

We now compare the effectiveness of hotspot pricing with respect to cordon pricing without requiring zero congestion. The main advantage of hotspot pricing is that it does not prevent vehicles from entering the Area-C zone, but it encourages that vehicles crossing conflicting hotspots avoid them, decongesting the hotspot and possibly its surrounding area. To compare the possible effects on the congestion of the city after the establishment of the two pricing schemes, we fix for both models the same revenue $\mathcal{P}$ and the same number of vehicles entering Area-C (which is equivalent to fixing the elasticity value). Specifically, $\mathcal{P}$ is the tax income received using the cordon pricing scheme (20M\euro{} reported in the literature).  We then compute the maximum number of vehicles that will avoid the taxed junctions, such that the remaining vehicles that accept to pay the tax produce total income equal to $\mathcal{P}$ (see appendix~\ref{app:maxCarsDistr}).
As in Eq.~\ref{eq:tax}, we consider that taxing junction $i$ at some price $c_i$ is enough to encourage a fraction $1-\phi_i$ of vehicles currently traversing $i$ to bypass it by choosing another non-taxed route. Under these conditions, we measure for every junction the accumulation of vehicles per minute (congestion), and compare their averaged distributions during weekdays for both models, see Fig.~\ref{fig:congDiffusionMilan}A. We observe that the median of the hotspot model is on average half the value of the cordon tax model, affording an improvement on the congestion of approximately 50\%.

Note that the number of vehicles within the Area-C is now different: while in the cordon taxing scheme it was reduced, in the hotspot pricing scheme it corresponds to the original flow of vehicles, but with a very different distribution over the city. Essentially, the distribution of congestion after the establishment of cordon tax scheme is not altered and it is still concentrated at the hotspots, as observed in real data. However, applying the hotspot scheme, the redistribution of vehicles is spread among neighboring junctions of the hotspots, which summarizes in less congested points even though the number of vehicles within Area-C is larger; remind we do not encourage vehicles to avoid entering the area. Graphical results about the congestion distribution in both cases are presented in Figs.~\ref{fig:congDiffusionMilan}B and~\ref{fig:congDiffusionMilan}C, respectively. We also report in Fig.~\ref{fig:times_same_revenue} the distribution of travel times without and with the taxing schemes. The hotspot pricing approach is able to yield significantly better travel times despite handling more vehicles in the Area-C than the cordon pricing.

\section{Potential effects of the hotspot pricing on air quality in Madrid}

\begin{figure}[tb]
	\centering
	\includegraphics[width=0.95\columnwidth]{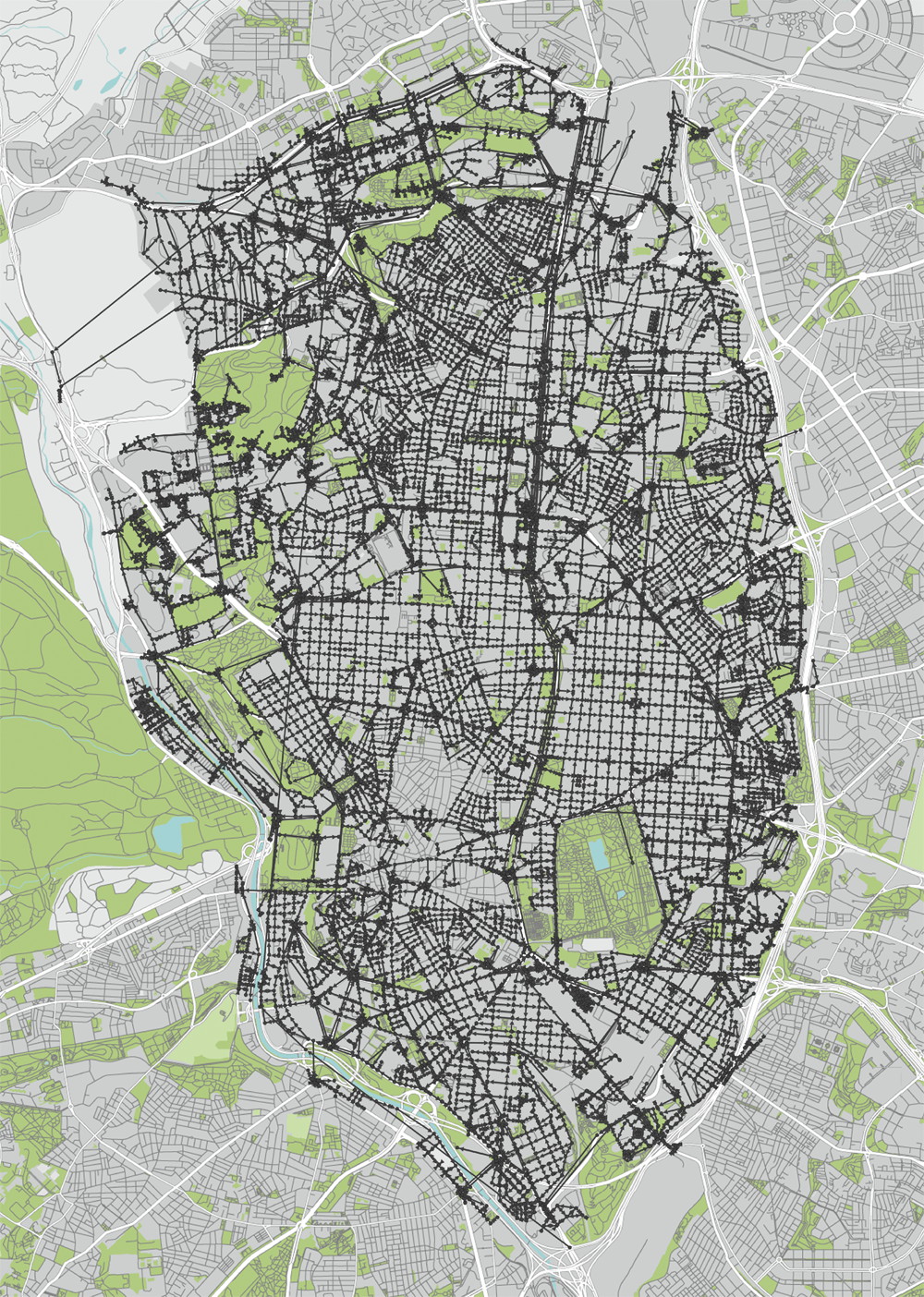}
	\caption{Topology of the road network inside the ring road M-30 of Madrid. Data gathered from Open Street Map (\url{http://www.openstreetmap.org}).}
	\label{fig:madrid_topology}
\end{figure}

\begin{figure}[tb]
  \begin{center}
	 \begin{tabular}{c}
		\includegraphics[width=.95\columnwidth]{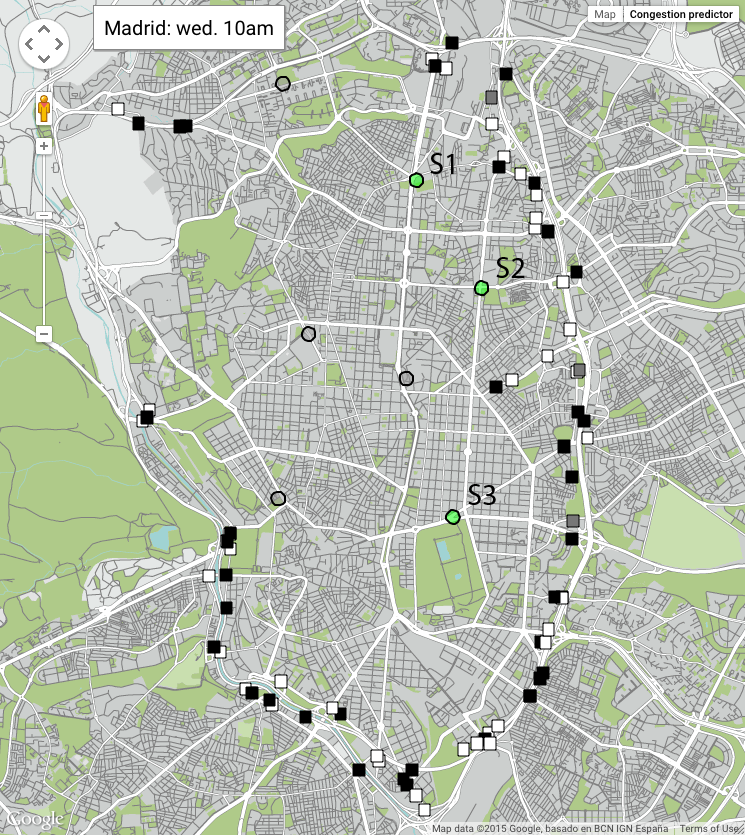}
 	 \end{tabular}
	\end{center}
	\caption{The image shows the manually selected~108 ingoing and outgoing gateways of Madrid and the~7 air quality stations of type ``Urbana tr\'afico'' within the selected area. City ingoing and outgoing gateways are symbolized with white and black squares respectively. Gray squares indicate gateways where vehicles travel in both directions. Air quality stations are represented by circles of 100~meter radius, green circles are stations for which the hotspot pricing scheme is expected to decrease the sensed $\mbox{NO}_2$ levels.}
	\label{fig:entryExitPointsAndSensingStations}
\end{figure}

The previous results show the hotspot pricing scheme could be a good alternative for managing traffic congestion in cities and this will probability have effects in their air quality. To give some hints of these effects, we have analyzed also the potential impact on the air quality of the city of Madrid (Spain) with a supposed establishment of the hotspot pricing scheme. Madrid city center is one of the most polluted areas in Spain, to the point that the Spanish government is pushing the city of Madrid to apply an urban tax to reduce pollution. Madrid is also the city of Spain where drivers waste more time in congestion (with an INRIX index of 10.8), followed by Bilbao (10.2) and Barcelona (8.6). The city plan of mobility includes the definition of a series of restricted traffic areas in the near future. To obtain the expected benefit of the hotspot pricing, we gathered data of the city topology and real traffic and pollution from Open Street Map and Open Data Madrid (\url{http://datos.madrid.es/portal/site/egob/}) respectively. Madrid open data portal provides the necessary information to obtain the expected contribution of cars to the overall city pollution. Madrid does not have any pricing zone so we apply the analysis to the zone delimited by the Madrid ring road M-30. The city topology and the entry and exit gateways have been obtained using Open Street Map. For the city topology, we have followed an equivalent procedure to Milan. The resulting topology can be seen in Fig.~\ref{fig:madrid_topology}. For the ingoing and outgoing gateways, we have manually selected the 108 roads crossing the M-30. Each cross point was selected to be an ingoing or outgoing gateway, depending on the road direction. Then, we have gathered traffic count point locations from the Madrid Open Data portal\footnote{\url{http://datos.madrid.es/portal/site/egob/}} and have assigned each of the 108 gates to the closer traffic count point. Figure~\ref{fig:entryExitPointsAndSensingStations} shows the gateway locations.

\begin{figure*}[tb]
  \begin{center}
	 \begin{tabular}{ll}
		{\bf A} & {\bf B} \\
		\includegraphics[width=0.45\textwidth]{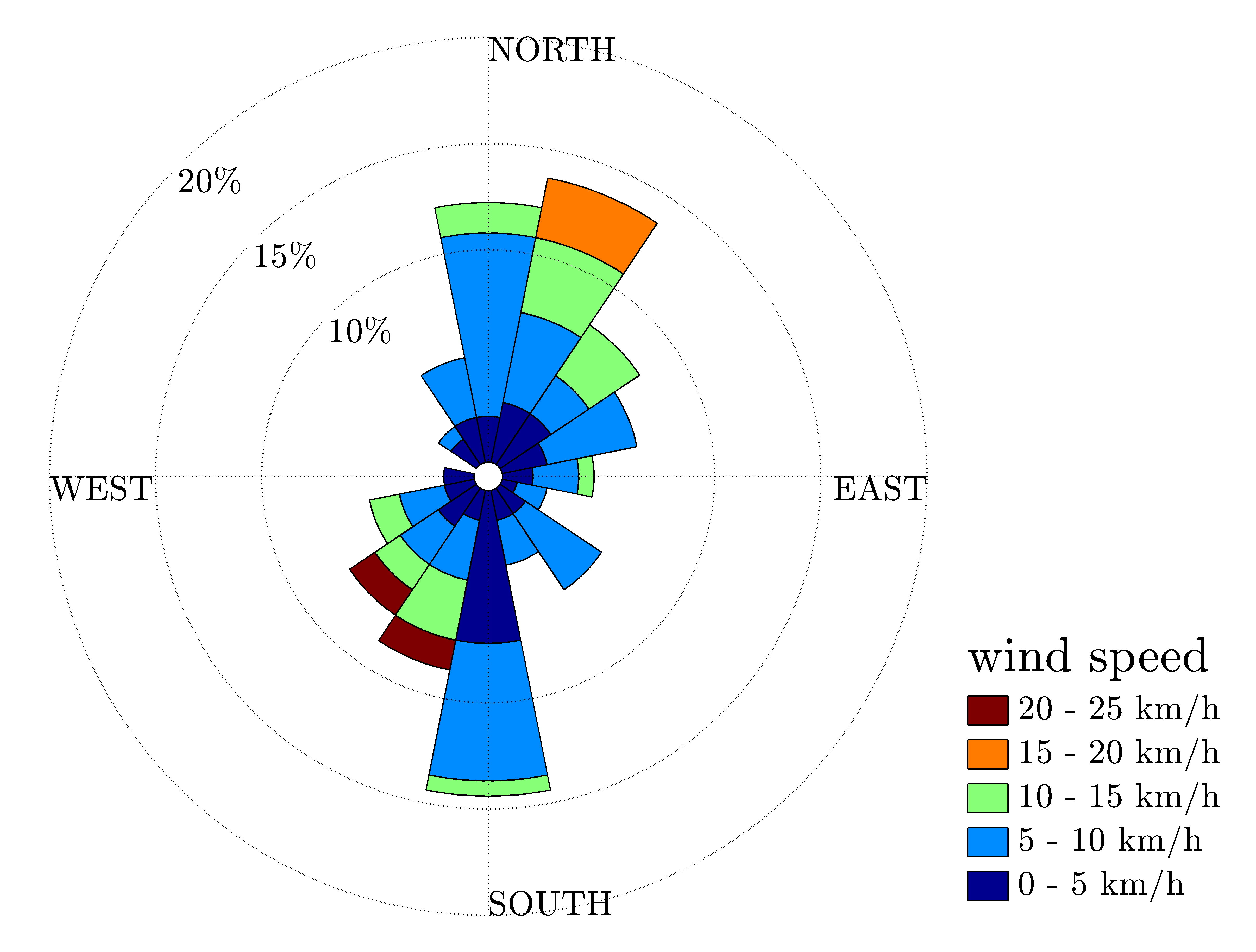} &
		\includegraphics[width=.45\textwidth]{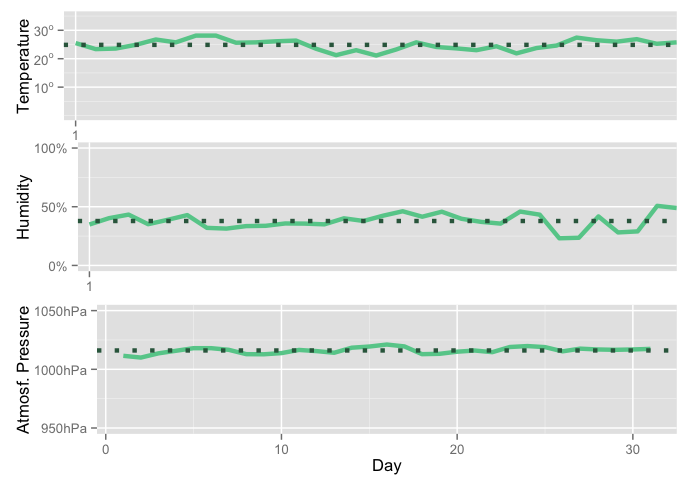}\\
		{\bf C} \\
		\multicolumn{2}{c}{\includegraphics[width=0.90\textwidth]{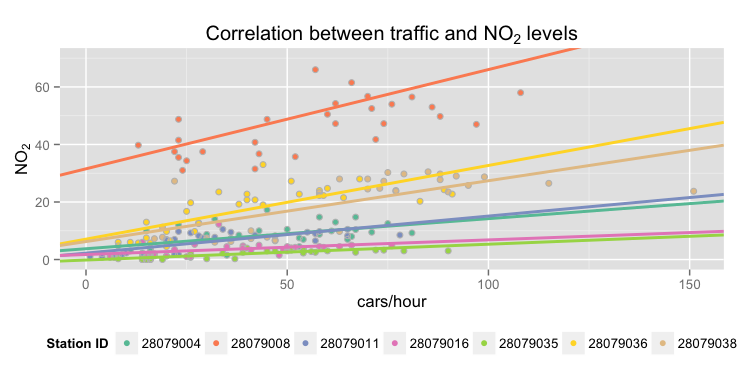}}\\
 	 \end{tabular}
	\end{center}
	\caption{Panels ({\bf A}) and ({\bf B}) show the main meteorological data (wind speed, preassure, temperature and humidity) of Madrid during August 2014. ({\bf C}) Scatter plots of vehicle flux with respect to $\mbox{NO}_2$ concentration, and the corresponding linear fits for each air quality stations.}
	\label{fig:metMadrid}
\end{figure*}

To obtain the contribution of each car to the overall pollution, we have taken data of the pollution and traffic levels of August 2014, which is one of the most stable months in terms of meteorology. Air pollution levels of Madrid\footnote{\url{http://www.mambiente.munimadrid.es/opencms/opencms/calaire}} have been obtained for each sensing station type ``Urbana tr\'afico'', i.e.\ stations located near main roads. See Fig.~\ref{fig:entryExitPointsAndSensingStations} for the location of the sensing stations. We have focussed in the $\mbox{NO}_2$ levels since it is known that, in Madrid, approximately 77\% of the $\mbox{NO}_2$ concentration comes from vehicles\footnote{\url{http://www.mambiente.munimadrid.es/opencms/opencms/calaire/ContaAtmosferica/portadilla.html}}. Then, we have accumulated the flux of vehicles of all count points within a distance of $100$~meters to each sensing station. With this information we have built a linear model to predict the contribution of vehicles to the pollution sensed by every station. The August meteorological data, the scatter plots of vehicle flux with respect to $\mbox{NO}_2$ concentration, and the linear fits are shown in Fig.~\ref{fig:metMadrid}. We have obtained an average slope of $0.16$ meaning that, in average, each car per hour contributes to $0.16 \mu g/m^3$ of $\mbox{NO}_2$ to the sensing station. The results are in perfect agreement with similar studies \cite{Raducan2009PolutansBucarest}.

\begin{figure*}[tb]
  \begin{center}
	 \begin{tabular}{ll}
	 	{\bf A} & {\bf B} \\
		\includegraphics[width=.40\textwidth]{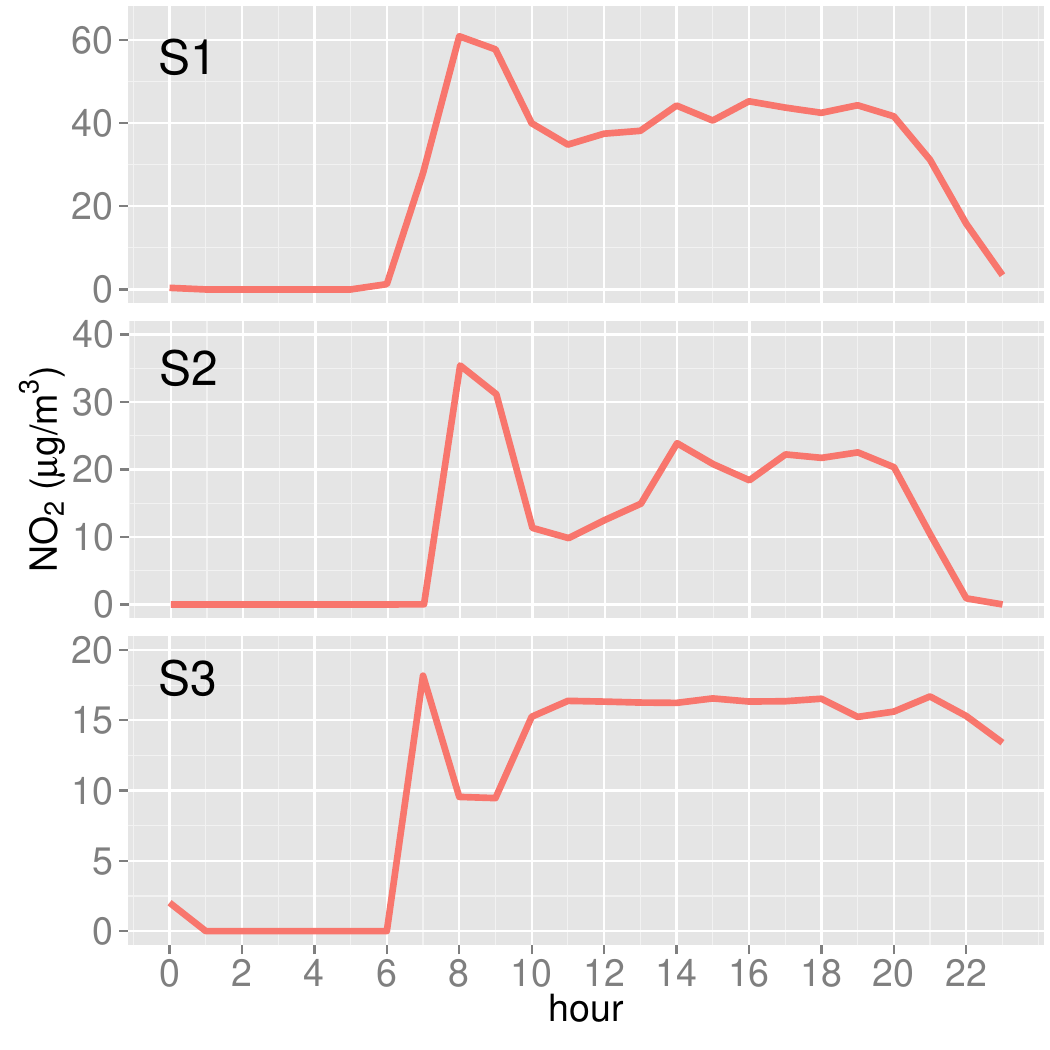}&
		\includegraphics[width=.40\textwidth]{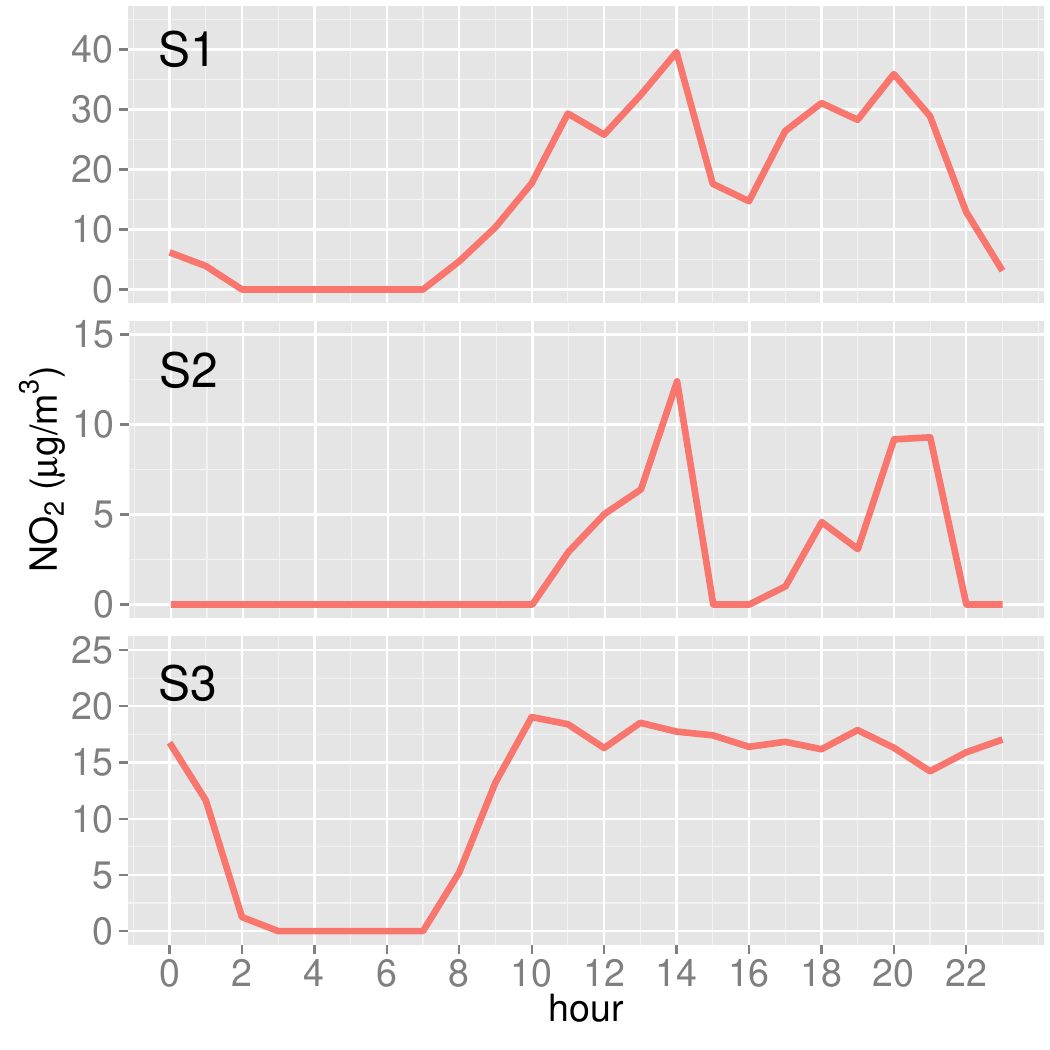} \\
	 	{\bf C} & {\bf D} \\
		\includegraphics[width=.40\textwidth]{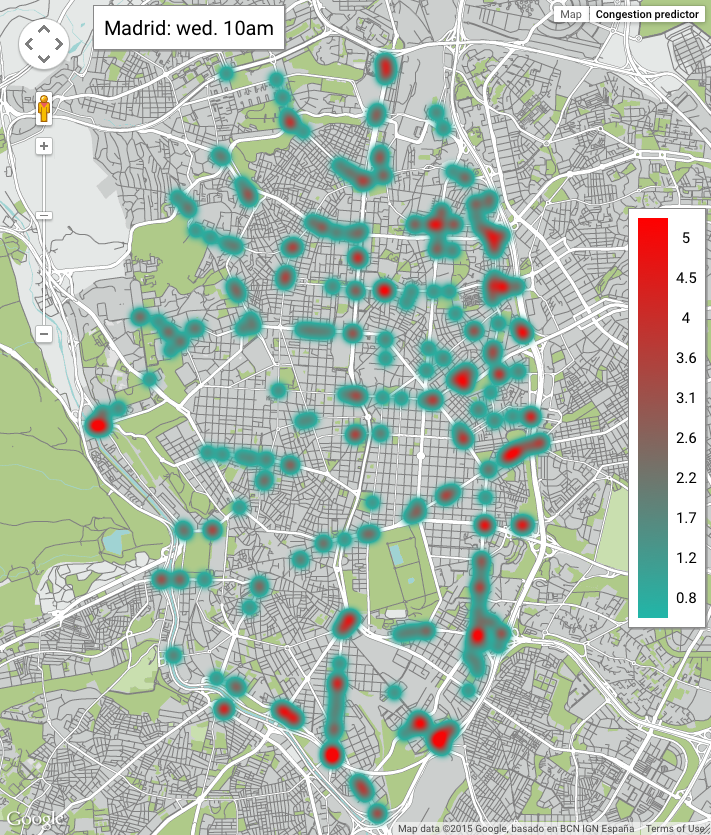}&
		\includegraphics[width=.40\textwidth]{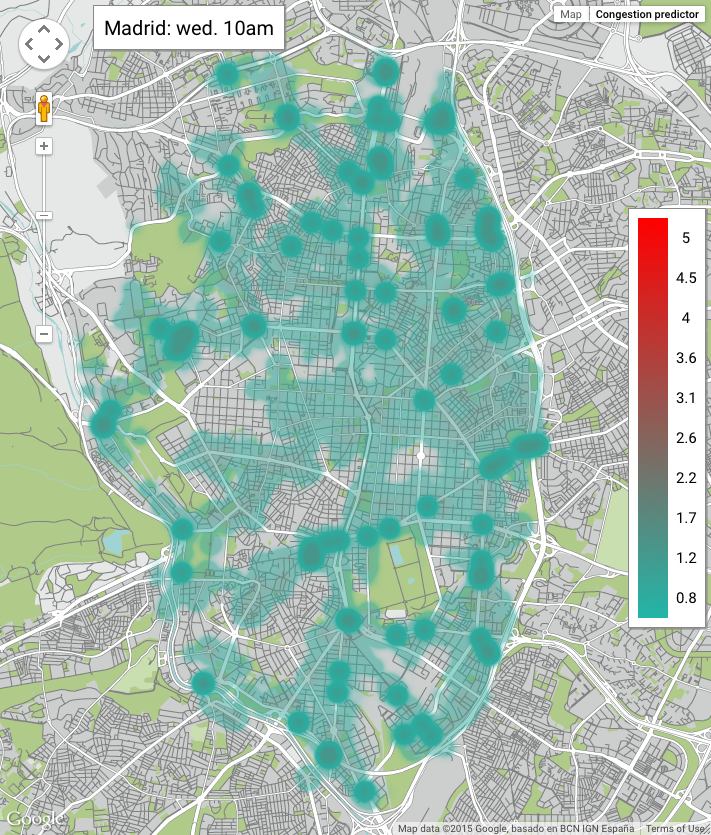}
 	 \end{tabular}
	\end{center}
	\caption{Panels ({\bf A}) and ({\bf B}) show the expected decrement of $\mbox{NO}_2$ concentration for~S1, S2 and~S3 stations. Plots consider data from Monday to Friday ({\bf A}) and weekend ({\bf B}) and are grouped by hour of the day. Circles colored with green in Fig.~\ref{fig:entryExitPointsAndSensingStations} indicate the locations of S1, S2 and S3 stations. ({\bf C}) Map of Madrid showing the ratio between incoming and outgoing vehicles for each congested junction before the establishment of the hotspot pricing scheme. Junctions with a ratio greater than~1 are congested since they receive more vehicles than the ones they can route. ({\bf D}) Ratio between incoming and outgoing vehicles for the same junctions shown in ({\bf C}) after the establishment of the hotspot pricing scheme.}
	\label{fig:exp_pollution_red_madrid_map}
\end{figure*}

To simulate the traffic dynamics of Madrid, we have analyzed the traffic data of October 2014 and obtained the average flow of vehicles per day and hour of the week. To analyze the possible effects on the air pollution after the introduction of the hotspot pricing scheme, we have computed the expected traffic of each junction of the city and then the expected traffic after the application of the hotspot pricing scheme. With the difference of vehicle flux in each junction we have computed the maximum possible reduction in $\mbox{NO}_2$ concentration. This maximum reduction assumes that the hotspot pricing motivates vehicles to bypass congested junctions, choosing other routes outside the 100~meter radius of the sensing station. Results are shown in Fig.~\ref{fig:exp_pollution_red_madrid_map}. After the hotspot pricing is applied, we observe a reduction of the level of $\mbox{NO}_2$ in 3 out of 7 air quality stations inside the ring road M-30 (see Fig.~\ref{fig:entryExitPointsAndSensingStations} for the location of the sensing stations). Clearly, this represents a local reduction and we cannot claim it implies a global pollution reduction. Panels~(A) and~(B) show the expected reduction per hour of the day for weekdays and weekends respectively. The larger reductions are observed in the morning rush hour, approximately from~7 to~10 on the weekdays and from~9 to~14 for the weekends. Panels (C) and (D) show the expected scenario before and after the the hotspot pricing. The overall amount of $NO_2$ is not reduced since the amount of cars in both scenarios is exactly the same. However, as expected, congestion, which was strongly centralized in several junctions, spreads and ameliorates within neighboring junctions.

\section{Conclusions}

Summarizing, traffic congestion is a common and open problem whose negative impacts range from wasted time and energy, unpredictable travel delays, and an uncontrolled increase of air pollution. Here, we have presented a hotspot pricing scheme, characterized by the application of local taxing policies instead of area taxing. The results are competitive reducing congestion and consequently pollution. We have shown two real case scenarios computing specific values of congestion and expected revenues.
These results pave the way to a new generation of physical models of traffic on networks within the congestion regime, that could be very valuable to assess and test new traffic taxing policies on urban areas in a computer simulated scenario.

\section{acknowledgements}
This work has been supported by Ministerio de Econom\'{\i}a y Competitividad (Grant FIS2015-71582-C2-1) and European Comission FET-Proactive Projects MULTIPLEX (Grant 317532). A.A.~also acknowledges partial financial support from the ICREA Academia and the James S. McDonnell Foundation.


\appendix
\section{Appendix}
\subsection{Optimal traffic redistribution given fixed revenue}\label{app:maxCarsDistr}

We want to compute the maximum fraction of vehicles, $1-\phi_i$, that will avoid the junctions with the hotspot pricing by fixing the overall tax income for the city $\mathcal{P}$. This may happen when local authorities want to fix the economic effort of the drivers to improve the traffic conditions. This is equivalent to the following minimization problem:
\begin{equation}
  \min_{\{\phi_i\}}\left(\sum_i \phi_i \sigma_i \right)\ \ \mbox{s.t.}\ \ \ \sum_i \phi_i \sigma_i c_i =\frac{c_{0}}{\phi_{0}^{1/\mu}} \sum_i \phi_i^{(\mu + 1)/\mu} \sigma_i = \mathcal{P}\,,
  \label{maxCarsDistribution}
\end{equation}
where $\sigma_i$ is the amount of cars junction $i$ receives before the taxing, $c_{0}$ is the initial price to obtain a reduction of $\phi_{0}$, $\mu$ is the elasticity and $c_i$ is defined in eq.~\ref{eq:tax}. The linear problem stand for the remaining cars that will cross the congested junctions after the taxing is applied and the restriction stands for the overall income produced by those cars.

We solve the minimisation problem using Lagrange multipliers. The objective function is
\begin{equation}
  L(\phi_i,\lambda) = \sum_i \phi_i \sigma_i - \lambda \( \sum_i {\phi_i}^{k_2}{\sigma_i} - k_1 \)\,,
\end{equation}
where $k_1 = \frac{\mathcal{P} \phi_{0}^{1/\mu}}{c_0}$ and $k_2 = \frac{\mu+1}{\mu}$. Setting the gradient $\nabla L\(\{\phi_i\},\lambda\) = 0$ we have:
\begin{align}
	\frac{\partial L}{\partial \phi_j} &= \sigma_j - \lambda k_2 {\phi_j}^{k_2-1}\sigma_j = 0 &&\implies \phi_j = \(\frac{1}{\lambda k_2}\)^{\frac{1}{k_2-1}} \label{phi}\,, \\
	\frac{\partial L}{\partial \lambda} &= \sum_i {\phi_i}^{k_2}\sigma_i - k_1 = 0 &&\implies \sum_i {\phi_i}^{k_2} \sigma_i = k_1 \label{lambda}\,.
\end{align}
From Eq.~(\ref{phi}) we see that all the $\phi_j$ are equal, i.e.\ independent of the node. Substituting Eq.~(\ref{phi}) into Eq.~(\ref{lambda}) we can obtain $\lambda$. Specifically,
\begin{eqnarray}
	\(\frac{1}{k_2\lambda}\)^{\frac{k_2}{k_2-1}} \sum_i {\sigma_i} = k_1\ \implies \
		\lambda = {k_2}^{-1} \({\frac{k_1}{\sum_i \sigma_i}}\)^{\frac{1-k_2}{k_2}}\,,
\end{eqnarray}
which yields an homogeneous fraction
\begin{eqnarray}
	\phi= \left(\frac{\mathcal{P} {\phi_0}^{1/\mu}}{c_0 \sum_i \sigma_i}\right)^{\frac{\mu}{\mu+1}}\,.
	\label{eq:redutionFixedP}
\end{eqnarray}

The local reduction to be applied is given by Eq.~\ref{eq:redutionFixedP}, and the tax to apply to every congested junction is
\begin{eqnarray}
	c = \({\frac{c_0^{\mu}\mathcal{P}}{\phi_0\sum_i \sigma_i}}\)^{1/(\mu+1)}\,.
	\label{eq:costFixedP}
\end{eqnarray}

\clearpage


\end{document}